\documentclass[12pt,preprint]{aastex}

\newcommand{\ie}{i.e.,\ }
\newcommand{\eg}{e.g.,\ }
\newcommand{\cf}{c.f.,\ }
\newcommand{\etal}{et~al.\ }
\newcommand{\ltsima}{$\; \buildrel < \over \sim \;$}
\newcommand{\simlt}{\lower.5ex\hbox{\ltsima}}
\newcommand{\gtsima}{$\; \buildrel > \over \sim \;$}
\newcommand{\simgt}{\lower.5ex\hbox{\gtsima}}

\newcommand{\magsec}{mag~arsec$^{-2}$} 

\def\a1413{Abell~1413}
\def\mkw7{MKW~7}

\def\rquart{$r^{1/4}$~}
\def\parcmin{{\tt '}\mskip -6.0mu.\,}
\def\parcsec{{\tt ''}\mskip -6.0mu.\,}
\def\muv{$\mu_{\mbox{v}}$}

\begin{document}
 
\title{Deep CCD Surface Photometry of Galaxy Clusters I: Methods and
Initial Studies of Intracluster Starlight}

\author{John J. Feldmeier\altaffilmark{1},  J. Christopher 
Mihos\altaffilmark{2}, Heather L. Morrison\altaffilmark{1,2}, 
Steven A. Rodney, Paul Harding}
\email{johnf@eor.astr.cwru.edu, hos@burro.astr.cwru.edu, 
heather@vegemite.astr.cwru.edu, sar9@smaug.astr.cwru.edu, 
harding@billabong.astr.cwru.edu}

\affil{Department of Astronomy, Case Western Reserve University,
10900 Euclid Ave, Cleveland, OH 44106}

\altaffiltext{1}{Visiting Astronomer, Kitt Peak National Observatory, 
National Optical Astronomy Observatory, which is operated by 
the Association of Universities for Research in Astronomy, Inc. 
(AURA) under cooperative agreement with the National Science Foundation.}

\altaffiltext{2}{Cottrell Scholar of Research Corporation and 
NSF CAREER fellow; Also Department of Physics, Case Western Reserve 
University.}

\begin{abstract}

We report the initial results of a deep imaging survey of
galaxy clusters.  The primary goals of this survey are to quantify 
the amount of intracluster light as a function of cluster properties, 
and to quantify the frequency of tidal debris.  We outline the
techniques needed to perform such a survey, and we report 
findings for the first two galaxy clusters in the survey: 
\a1413, and \mkw7.  These clusters vary greatly in richness and
structure.  We show that our surface photometry reliably reaches
to a surface brightness of \muv = 26.5 \magsec.  We find that both clusters
show clear excesses over a best-fitting \rquart profile: this was
expected for \a1413, but not for \mkw7.  Both clusters also show
evidence of tidal debris in the form of plumes and arc-like 
structures, but no long tidal arcs were detected.  We also find
that the central cD galaxy in \a1413 is flattened at large radii, 
with an ellipticity of $\approx 0.8$, the largest measured ellipticity 
of any cD galaxy to date.  
\end{abstract}
 
\keywords{galaxies: clusters: general --- 
galaxies: clusters: individual (Abell~1413, MKW~7) --- galaxies: 
interactions -- galaxies: kinematics and dynamics}
 
\section{Introduction}

The concept of intracluster starlight was first proposed by \citet{zwicky1951},
who claimed to detect excess light between the galaxies of the Coma
cluster.  Follow-up photographic searches for intracluster luminosity in
Coma and other rich clusters \citep[\eg][see V\'ichez-G\'omez 1999 \&
Feldmeier 2000 for reviews]{wel1971,mel1977} produced mixed 
results, and it was not until the advent 
of CCDs that more precise estimates of the amount of intracluster starlight 
were made \citep[\eg][]{uson1991b,vg1994,bern1995,gon2000}. 
These observations are extremely difficult to perform and interpret 
due to the low surface brightness of the phenomenon:  typically, the surface
brightness of intracluster light is less than 1\% of the brightness of
the night sky.  Measurements of this luminosity must also 
contend with the problems presented by scattered light from 
nearby bright objects and the contribution of discrete sources.

Despite these difficulties, intracluster light (ICL) is of potentially
great interest to studies of galaxy and galaxy cluster evolution.
The dynamical evolution of cluster galaxies is complex, involving 
poorly understood processes such as 
galactic encounters, dark matter, cluster accretion, and tidal
stripping (cf.~Dressler 1984).  The ICL provides
a direct way to study these different mechanisms.
Various studies have suggested that anywhere between 10\% and 70\%
of a cluster's total luminosity may be contained in the ICL
(Richstone \& Malumuth 1983; Miller 1983), with a strong dependence
on the dynamical state of the cluster.
The properties of the ICL may also be sensitive to the distribution 
of dark matter in
cluster galaxies. Simulations have shown that the structure of dark
matter halos in galaxies plays a central role in the formation and
evolution of tidal debris \citep{dub1996,dub1999}.
If cluster galaxy dark halos are tidally truncated at small radii
\citep[\eg][]{whitmore1988}, stripped material can be more
easily unbound from the galaxies and end up being distributed smoothly
throughout the cluster. Conversely, if cluster galaxy halos survive,
some tidally stripped material may remain bound to these galaxies, 
leaving them embedded in very low surface brightness ``cocoons.''  The ICL
may act as a sensitive probe of the mechanics of tidal stripping, the
distribution of dark matter around galaxies, and cluster evolution in
general.

Recently, much progress has been made in the study of intracluster
starlight on numerous fronts.  Individual intracluster stars, namely
planetary nebulae detected from the ground and red
giants detected using {\sl HST}, have been discovered in the Virgo and 
Fornax clusters \citep{1996arna,t1997,mend1997,m87ipn,ipn1,ftv1998,
phd2000,durr2002}.
Although some of the intracluster planetary 
candidates were later found to be background objects 
\citep{kud2000,blank2002},
most are bona-fide intracluster planetary nebulae 
\citep{freeman2000,blank2002}.  There is also some evidence for
intracluster supernovae, though the results here are more 
tentative, and the rate may be significantly smaller than that
seen in galaxies \citep{smith1981,galyam2000,tyson2002}.  
These individual intracluster stars give the promise of studying 
in detail the kinematics, metallicity, and age of the intracluster 
stellar population in nearby galaxy clusters.

Another area of progress is the advent of modern numerical studies
of the dynamical evolution of galaxies in clusters.  High-resolution N-body
simulations now have the ability to follow hundreds of cluster galaxies
interacting within a cosmological context \citep{harass,dub1998}.  These
high-quality simulations finally allow testable predictions of the
production and properties of intracluster starlight 
\citep{mlk1998,dub2000,korch2001}.  When combined with earlier
theoretical studies, \citep{gal1972,mer1983,rich1983,mil1983,mer1984} 
there is now a growing theoretical framework in which to interpret 
observations of the ICL.

Recently, another aspect of intracluster starlight has been 
discovered: tidal debris arcs.  These features are large ($\sim$ 100 kpc),
low-surface brightness (\muv $\sim 26$ \magsec) 
arc-like structures found in nearby galaxy clusters, and are not
due to gravitational lensing.  These arcs have been found in the Coma
and Centaurus clusters \citep{tren1998,gregg1998,cr2000}.  It has been
proposed that these arcs are due to tidal interactions between
galaxies and the cluster's gravitational potential \citep{harass}.
Since several of these debris arcs were found by chance, it is
plausible that they might be present in other galaxy clusters.  The
scientific potential for these arcs is exciting.  By observing the
morphology, and -- in the future -- kinematics of these stellar
streams, much can be learned about the orbits of the infalling galaxies, 
and the gravitational potential of the galaxy cluster \citep[\ie][]{cr2000}.

Although the presence of intracluster stars has been clearly 
demonstrated, there is little information on how the amount and 
distribution of intracluster starlight varies with the 
properties of the cluster it inhabits.  We do not yet have a global 
picture of the prevalence of the ICL in galaxy clusters, nor of 
the information it contains about the dynamical state of clusters, 
both of which are crucial in developing more advanced models of 
cluster evolution.  Studies of individual intracluster stars are
invaluable in nearby clusters, but are flux-limited
and so cannot probe the evolution of intracluster light to
higher redshift.  Finally, there is little data on how common tidal
debris arcs might be in galaxy clusters.  Currently, the majority of 
tidal debris arcs discovered have been found in the Coma cluster 
\citep{tren1998,gregg1998,cr2000}.  
The Coma cluster is well known to be unusually rich 
\citep[\cf][]{dress1984}, and it is possible that 
it might contain an unusually large number of tidal debris arcs.

Finally, another interesting facet of the ICL in clusters is the nature 
of cD envelopes. First identified in deep photographic imaging of clusters,
cD galaxies are characterized by an excess of diffuse light (compared
to an \rquart\ law) at large radius. The origin of cD envelopes
remains unclear: are they formed in the initial stages of cluster
collapse, or later, as galaxies continue to fall in the cluster and 
become tidally stripped? The detailed light distribution of cD envelopes
may hold clues to the answer. The statistical mechanics of violent
relaxation naturally produces \rquart-like profiles \citep{lbell1966};
if cD envelopes form during cluster collapse, they should show
such a profile. On the other hand, continued accretion and
stripping of infalling galaxies need not produce an \rquart\ profile,
as the distribution of stripped light will be more sensitive to the
orbital energy and angular momentum of the infalling galaxies.
While the characterization of cD envelopes as an excess of light 
over an \rquart\ profile would seem to argue for a stripping origin,
most studies of cD galaxies have used older photographic data,
with very uncertain flat fielding characteristics. Newer work using
CCD imaging has shown at least one case where a cD galaxy may
in fact be well characterized by a pure \rquart\ law \citep{gon2000}.
In light of this result, revisiting the question
of the structure of cD galaxies using deep CCD imaging may shed new
light on the origin of the cD envelopes and the evolution of galaxy
clusters.

To address these questions, we have begun deep imaging of a sample of
galaxy clusters to quantify the structure of
ICL.  Although quantitative surface photometry several magnitudes
below sky is an extremely challenging task, over the past decade 
the necessary CCD imaging techniques have been developed and
carried out on both  galaxies
\citep[\eg][]{sb1994,sb1997,sb1999,zheng1999} and galaxy clusters
\citep[\eg][]{tyson1998,gon2000}. Using these techniques, our plan
is to image galaxy clusters which differ in richness, concentration,
and sub-structure to quantify how the ICL changes as a function
of environment. In conjunction with increasingly sophisticated
models of cluster galaxy evolution, such observations can
provide constraints on the evolution of clusters and cluster
galaxies, the formation of the ICL, and the distribution of
dark matter in cluster galaxies. In this paper, we describe in
detail our imaging techniques, and show results from the
first two clusters surveyed.

\section{Selection Criteria}

Our program is aimed at studying the ICL in clusters possessing a variety of
structural properties, in order to probe the relationship between the 
ICL and cluster environment.  Our initial sample will primarily focus on 
Abell \citep{aco} clusters of distance class 5--6 (z $\approx$
0.1--0.175) with differing richness and Bautz-Morgan classifications.  
The lower end of the redshift range is chosen such that the 
inner $\sim$ 0.75 Mpc of the cluster fits on the field-of-view 
of moderate size CCD detectors, allowing us to study 
the cluster as a whole without mosaicing, and permitting a 
reasonable amount of sky at the outer edge of the field 
for sky subtraction. The upper limit is
set so that $(1+z)^4$ surface brightness dimming is not prohibitive,
and also to prevent the angular size of any tidal arcs from being too
small.  For comparison purposes, we also observe nearby poor galaxy 
clusters from the MKW/AWM catalogs of poor clusters that appear to 
contain cD galaxies \citep{mkw,awm}.    
These clusters will provide a significantly different cluster 
environment in terms of density and dynamical interaction.

Scattered starlight is a crucial source of systematic error in our 
program.  Therefore, we must screen our candidate clusters carefully, 
making sure there are no bright stars in the CCD field or up to
several degrees nearby.  Because of the complex spatial nature of 
the scattered light distribution, we do our screening  
by manual inspection of the original Palomar Sky Survey 
(POSS) plates in the area around the target cluster.  
Approximately half of our candidate clusters are rejected by 
this process.  Once the cluster passes both the catalog 
criteria and scattered light tests, it is included as a potential 
target.

\section{Cluster Properties}
For the first targets of our survey, we chose two galaxy clusters with 
greatly differing properties. 
\a1413 ($\alpha$ = 11h 55m 22.5s, $\delta$ = +23\degr
22\arcmin 18\arcsec, J2000.0) is a galaxy cluster of richness class 
3 (richer than 95\% of the original Abell catalog), with a Bautz-Morgan type 
of I \citep[cD dominated;][]{leir1977} and a Rood-Sastry type of 
cD \citep{sr1987}.  Its central cD galaxy has been studied with
photographic surface photometry at large radii, and CCD surface
photometry at smaller radii 
\citep{oem1976,schombert1986,schombert1988,sgh1983,porter1991} 
numerous times, allowing us to compare our results with others.  
These earlier studies 
imply that the properties of A1413's cD halo are extremely impressive:
\citet{oem1976} found that the cD halo of \a1413 might extend as far as
24\arcmin~(2.4~$h^{-1}$~Mpc; H$_{0} = $ 100~$h$~km~s$^{-1}$~Mpc$^{-1}$) 
away from the center of the cluster, and \citet{morgan1965}
indicated that \a1413 might be the largest of all cD galaxies.
\citet{schombert1988}
studied \a1413 in detail, and found that the cD halo extended to
$\sim$ 500~kpc~h$^{-1}$, with
a large excess over the best-fitting de Vaucouleurs r$^{1/4}$ 
profile of the inner regions. \a1413 has a relatively high X-ray 
temperature \citep[8 keV;][]{white2000}, and has a Sunyaev-Zeldovich
decrement \citep{grainge1996}, confirming that the cluster 
is indeed massive.  Therefore, \a1413 
is a representative example of a rich cluster, albeit containing an
abnormally large cD galaxy.
 
\mkw7 (WBL 514 -- White \etal (1999); $\alpha$ = 14h 34m 00.9s, 
$\delta$ = +03\degr 46\arcmin 52\arcsec, 
J2000.0) is a poor galaxy cluster whose brightest cluster galaxy was found
to be cD-like in appearance from inspection of the Palomar Sky Survey \citep[]
[see Tonry 1987; Schombert 1992 for discussions of cD 
classification]{mkw}.  It has a richness class of 
-1 \citep{bahcall1980}, and in galaxy counts, is over a factor of 
eight poorer than \a1413.
Photographic surface photometry has been made of MKW~7's brightest
cluster galaxy \citep{morbey1983}, but these measurements are 
complicated by the presence of a $m_{\mbox{v}} \approx$~11.9 star 
within 21\arcsec~of the galaxy nucleus.
\citet{sidney1977} notes that the brightest cluster galaxy 
is embedded in a bright but asymmetrical envelope.
From galaxy density maps and redshift information \citet{beers1995} 
argue that MKW~7 is gravitationally bound to another poor cluster,  
MKW~8, which is within 1.5 h$^{-1}$~Mpc.  
\mkw7 has been detected in X-rays multiple times 
\citep[\eg][]{price1991}, but no gas temperature has 
yet been determined.  In comparison to \a1413, \mkw7 is a 
poorer, less dynamically evolved cluster.

\section{A Note on nomenclature}

We note that the term ``intracluster starlight'' has been applied in
many different ways in the literature. A dynamical definition might be
stars which are unbound from any individual cluster galaxy, yet bound
to the cluster as a whole. From an observational point of view, of
course, this definition is inaccessible without knowing the detailed
kinematics of the ICL and the total mass distribution in the
cluster. With deep imaging, the definition of ICL can only be made
based on the surface brightness distribution within the
cluster. Indeed, it is debatable whether cD envelopes should be
considered as ICL -- is the envelope a feature of the cD galaxy
itself, or are both the cD and the envelope simply material which has
collected at the bottom of the cluster potential well?  Uson, Boughn
\& Kuhn (1991) succinctly summarize the situation in their
observations of Abell 2029:

\begin{quotation}
``...Whether this diffuse light is called the cD envelope or diffuse 
intergalactic light is a matter of semantics; it is a diffuse component which
is distributed with elliptical symmetry about the center of the cluster 
potential...''
\end{quotation}

Based on surface photometry alone it is difficult to disentangle
luminosity associated with a cD envelope from that of the
extended ICL, and in fact such a distinction may not be well
motivated from a physical point of view. For the purposes of our
work, we will simply use the term ``intracluster light'' to
refer to the diffuse light in galaxy clusters, and address 
issues related to cD envelopes, diffuse arcs, and extended ICL
in the context of diffuse light as a whole.

\section{Observations and Reductions}

\subsection{Observations}
The data for \a1413 and \mkw7 were obtained over two photometric
nights during a four night run in 2000 April, using the 2.1m telescope 
at Kitt Peak National Observatory\footnote{Kitt Peak National 
Observatory is a division of NOAO, which is
operated by AURA, under cooperative agreement with the National Science
Foundation.}.  The images were taken using a 2048 x 2048 
Tektronix CCD (T2KA).  With this setup, the field of view is 
10.4 arcmin$^{2}$, with each pixel imaging 0.305\arcsec of sky.  The
gain was set at the default value of 3.6 e$^{-}$~ADU$^{-1}$ and the  
readout noise was 4~e$^{-}$ (1.1 ADU).  All exposures were made 
through a Washington {\it M} filter, which is similar 
to Johnson {\it V} but is slightly bluer in mean wavelength and
therefore contains fewer airglow emission lines (see Figure 1).  
These airglow lines, produced in the upper atmosphere by a variety of 
mechanisms 
\citep{roach1973}, are a significant source of sky background, and are
well known to be variable over timescales of minutes
\citep[\eg][]{pil1989,kris1997,sb1997}.  Therefore, to reduce the sky 
background, and to simplify the process of sky subtraction 
and flat-fielding, we chose the Washington {\it M} filter for our
observations.  We transformed these observations to Johnson {\it V} 
(\S 5.4), and unless otherwise stated, all surface brightnesses are
given in {\it V} magnitudes.  

We began the telescope run by preparing the detector, telescope
and filter to reduce the amount of scattered light, which sets the
ultimate limit to our surface photometry.  
We first placed a black cardboard mask over the detector's dewar 
window in order to reduce scattered light
from the mounting hardware surrounding the CCD.  These
metallic components are highly reflective, and a clear
difference in the amount of scattered light is readily apparent.  
We next took pin-hole images of the telescope pupil 
to search for other sources of scattered light \citep{grundahl1996},
and baffled any such areas with black cloth.  

An accurate flat-field is critical to the success of our program. 
As mentioned in \S 1, we are interested in recovering a signal that
is less than 1\% of the sky background.  Our flat field must be at
least a few times more accurate than this 1\% value for our data to
be meaningful.  For this reason, dome flat fields cannot be used due to
possible scattered light, differing pupil illuminations, and intrinsic
color differences.  For similar reasons, twilight flats are
also inadequate for our purposes.  Therefore, dark sky flats are a 
neccesity, and we performed the observations in the manner described 
by Morrison \etal(1997).  Half of the time was used observing the 
galaxy clusters, and the other half was used to obtain dark sky flats.  
The dark sky flat images were taken at pre-determined areas away 
from bright stars at approximately the same hour angle and 
declination as the cluster images.  Over the course of the 
observing run, a total of nine images were taken of \a1413, 
12 images were taken of \mkw7, and 23 dark sky flats were 
obtained.  For each of the cluster and sky images, the exposure 
time was 900~s.  

\subsection{Correction for non-linearity:}
After our run, we were made aware of the presence 
of non-linearity in the T2KA detector
by K. Stanek (reported in Mochejska \etal 2001).  Figure 2
shows the comparison of stellar photometry derived from a 
60s test exposure of \mkw7 and a median-combined exposure of 900s, both
reduced in the standard manner.  Although the scatter is large, 
a clear non-linearity is present in the data.  We fit the residuals 
in magnitude with a least-squares linear model, and found
a residual slope of $0.0095 \pm 0.0008$ magnitudes per magnitude.  
This is consistent with the
measurements of \citet{moch2001} for their test photometry
of NGC~7789 (see their figure 3).  Since the non-linear behavior is 
identical to that seen in the data presented in Mochejska \etal (2001), 
the data was taken with the same instrument only 5-6 
months apart, and the Mochesjska \etal data better constrains the
effect, we adopt identical corrections for non-linearity: 

\begin{equation}
I_{e} = I_{i} \cdot (c_{1} + c_{2} \cdot \frac{I_{i}}{32767} + 
c_{3} \cdot (\frac{I_{i}}{32767})^{2})
\end{equation}  
where $I_{i}$ is the measured intensity, and $I_{e}$ is the corrected
intensity in ADU.  The constants derived by Mochejska \etal (2001)
for a gain of 3.6 e$^{-}$~ADU$^{-1}$ are:

\begin{equation}
c_{1} = 0.983282, c_{2} = -0.0765595, c_{3} = 0.0252555
\end{equation}  
Following Mochejska \etal (2001), we used 
the IRLINCOR task within IRAF\footnote{
IRAF is distributed by the National Optical Astronomy Observatories,
which are operated by the Association of Universities for Research
in Astronomy, Inc., under cooperative agreement with the National
Science Foundation.}, to correct the data for non-linearity
after overscan removal and bias subtraction. 

For two reasons, the non-linearity is less of a problem for
our project than it might first seem.  First, since our sky flats have
exactly the same exposure time as our data, they have the identical
non-linearity inherent in the exposures, so any difficulty in
flat-fielding is canceled out to first order.  Second, at low 
surface brightness levels, the error caused by any nonlinearity 
is significant, but relatively small.  Figure 3 shows the difference 
in magnitudes between the corrected, and non-corrected 
sky-subtracted surface brightness.
Nonetheless, any error in the correction for non-linearity adds a
source of error to our surface brightness estimates.  
To determine the amount of such error, we obtained the linearity data
from Mochejska \etal (2001) (kindly provided to us by B. Mochejska), 
and replicated the linearity fit.  We found the identical 
constants with the following 1-$\sigma$ errors on the parameters:
\begin{equation}
\sigma_{c_{1}} = 0.0052,   \sigma_{c_{2}} = 0.012,   \sigma_{c_{3}} = 0.0057
\end{equation}  
This uncertainty is added to our final error model (see \S 6.4).

\subsection{Flat Fielding}

After overscan removal and bias subtraction (done in the standard
manner), we constructed a ``master'' sky flat from the dark sky
images taken.  First, each individual sky flat was visually inspected
to ensure that no bright stars or scattered light patterns were
present in the image.  This is necessary because in some of our exposures of 
candidate galaxy clusters at the telescope, we found that a grid-like 
scattered light pattern appeared.  This
pattern was rotated 45 degrees from the CCD axes, and typically
covered an area of 370 by 350 pixels.  When the pattern did appear, 
its surface brightness varied, but it could be as large
as \muv $\approx 25.2$ \magsec.  
Three of the dark-sky images were found to have unacceptably large scattered 
light patterns, and were removed from the list, leaving 20 
dark-sky images to construct the dark sky flat.  Three other dark sky 
images also contain the scattered light pattern, but the amplitude 
of the pattern was so small in these cases (their presence was 
barely visible on the image) that they were left in the sample.  
No scattered light patterns were seen in any image 
of \a1413 and \mkw7. 

To construct the best possible master sky flat, we combined the
individual sky frames using a very accurate determination
of their modes.  We do this using the iterative procedure
described in Morrison \etal (1997).  We first pre-scaled the 
images by their mode, found from the IMSTAT task
within IRAF.  Prescaling is important because even with our 
relatively narrow {\it M} filter, the modal value of the 
sky images varies by up to 22\%, from maximum to minimum.  
Then we combined the individual, modal-divided sky images to make a
preliminary flat-field frame, using IRAF's IMCOMBINE task, with the
CCDCLIP algorithm, set to remove pixels which differed from 
the median by more than 2 $\sigma$.  

Each of the individual sky frames were then divided by this preliminary 
flat-field frame to reduce the width of the distribution of modal 
sky values, making rejection of outliers due to faint stars 
and stellar wings more accurate.  The flat-fielded 
sky frames were then averaged into 50 by 50 pixel bins, and 
a plane was then fit to the binned-up images 
using the IRAF task IMSURFIT.  This step is necessary because each 
individual sky frame has noticeable sky variations across the 
image due to a number of atmospheric effects such as airglow 
\citep{roach1973,wild1997,zheng1999}.  Figure 4 shows the 
binned-up images for each sky frame after they have been flat-fielded.  
Clear systematic sky variations can be easily seen in the data.  
After the individual planes were fit and normalized, the sky
frames were each divided by their normalized plane.  The modes were then
recalculated using our own software, and the entire procedure was 
repeated using the improved flat-field frame.  
The procedure was repeated until the calculated modes had converged 
(about 15 times in this case).  

The galaxy cluster images were then flat-fielded by this final flat,
and were registered using stars common to all frames and the IRAF 
tasks GEOMAP and GEOTRANS, using a 2nd order polynomial fit.  
A preliminary sky value was found for each cluster image by finding
the mode of two regions on each chip well away from the 
center of the cluster, and averaging the results.  
This sky value was then subtracted from each image.  
The median sky value for \a1413 was 886.0 ADU~pixel$^{-1}$ and 
932.7 ADU~pixel$^{-1}$ 
for \mkw7.  After applying the photometric zero point in (\S 5.4)
below, these correspond to \muv = 21.11 \magsec~and \muv = 21.05 \magsec, 
respectively.  Since the source of sky brightness is mostly  
within the earth's atmosphere, we remove our 0.17 mag~airmass$^{-1}$
extinction correction, and find that the average brightness of the
night sky at zenith was approximately \muv = 21.25 \magsec, in 
reasonable agreement with the solar maximum value of $21.287 \pm 0.048$ 
of \citet{kris1997}. 

With the overscan, bias-subtraction, flat-fielding and sky subtraction 
complete, we then combined the images together, using a 2$\sigma$ clipped 
median as before, and scaling for airmass.  The final images for 
\a1413 and \mkw7 are displayed in Figures 5 and 6.  The measured 
seeing (full-width at half-maximum) for the final combined images 
was $1\parcsec22$ for \a1413 and $1\parcsec37$ for \mkw7.

\subsection{Photometric Zero Point:}
The Landolt star fields SA~98, SA~107, \& SA~110 (Landolt 1992) 
were observed, giving us a total of 37 well-observed 
standard stars over a range of color and airmass.  
For the purposes of our analysis, we converted our Washington 
{\it M} exposures to {\it V} band magnitudes.  This transformation
is straightforward because all of the Washington {\it M} standard 
stars used in these observations are also Landolt (1992) {\it V} 
standards.  A photometric zero point of V = 21.09 $\pm 0.04$ mag 
arcsec$^{-2}$ (corresponding to 1~ADU~s$^{-1}$~pixel$^{-1}$, and 
assuming a (B-V) color of 1.0 ) was determined.  For a 900 s exposure 
this yields  V = 28.48 mag arcsec$^{-2}$ corresponding to 1~ADU per
pixel at unit airmass.  As our exposures were only taken in 
one filter, and we do not know the exact color of the intracluster
light, we cannot add a color correction term to our target photometry, 
but from the standard star observations, we estimate its magnitude 
as less than 0.1 mag, over the entire likely color range of our 
target objects (0.8 $\leq$ (B-V) $\leq$ 1.3).  The color term is 
reasonably well fit as a linear function of (B-V), with a slope of 
0.2 magnitudes per magnitude of (B-V) color.

\section{Analysis and Results}

We adopt an approximate angular size distance 
to \a1413 and \mkw7 of 465 and 111 Mpc respectively, assuming 
redshifts of z=0.1427 for \a1413 \citep{sr1999} and z=0.0290 
for \mkw7 \citep{beers1995}, a Hubble constant, H$_{0} = 
$ 75~km~s$^{-1}$~Mpc$^{-1}$, and a cosmology of $\Omega_{\mbox{m}} =
0.3$, $\Omega_{\Lambda} = 0.7$.  At these small redshifts, these 
distances depend little on $\Omega$.  Given these assumed distances, 
1 arcsecond subtends $\approx$ 2.3 kpc in \a1413, and $\approx$ 
0.52 kpc in \mkw7.  The corresponding luminosity distance moduli, 
ignoring any K-corrections, are 39.0 for \a1413, and 35.4 for \mkw7.

\subsection{Masking}

In order to reach the faintest possible surface brightness levels of 
the cD galaxy + intracluster light, we must mask out all other sources --- 
both stars and galaxies --- in the frame.  We begin by creating a binary 
mask image where one indicates a good pixel and zero indicates a bad pixel.  
This has the advantage of allowing us to visually compare 
our mask image at any point in the construction process by simply 
multiplying the mask by the data image, and displaying the results. 

We first begin by masking out the stars in each image.  Since we are 
concerned with very low surface brightness, we must determine 
the point spread function (PSF) out to very large radii.  
Using the DAOPHOT (Stetson 1987) package, we detected all of the stars in the 
frame down to a signal-to-noise of three, and used a subset 
of bright stars to first determine the PSF 
out to a radius of 20 pixels.  We then used this preliminary 
PSF to mask out all of the stars and small galaxies around two bright
saturated stars in our \a1413 data.  Saturated stars have much higher
signal-to-noise in the wings on the PSF, which are our primary concern.  
Other sources, such as resolved galaxies and stellar 
diffraction spikes, were removed manually.  Then the 
unmasked pixels from the two saturated stars were averaged
in radial annuli, and joined to the preliminary PSF (which measures
the inner core of the star more accurately).  The final radial profile
is displayed in Figure 7.  Using this large-radius 
PSF, and the list of stars found by DAOPHOT, we masked all 
stellar sources in the frame out to a radius where the
magnitude-scaled PSF was 1 ADU above the sky value.  

Next, we must mask out all of the galaxies in each cluster, 
excluding the central cD.  Unresolved background galaxies have been  
treated as point sources, and have already
been masked by the DAOPHOT procedure above, but many resolved sources
remain in both clusters.  We chose to mask out the galaxies using the
segmentation image from the SExtractor software package 
\citep[V2.2.1;][]{sex1996}.
Again, since we must mask down to very low surface brightness levels,
the SExtractor detection parameters are set for faint surface
brightness levels.  After experimentation, we adopted a 
minimum detection threshold 
of 4 pixels that were 0.6 $\sigma$ above the local sky background.  
This corresponds to 3.3 ADU in 
\a1413, and 2.7 ADU in \mkw7.  Assuming Gaussian statistics, 
the probability of a false SExtractor 
detection at these low-light levels is $5.6 \times 10^{-3}$ per four
pixel block.  This is uncomfortably high, and allows for the
possibility of ``over-masking'' our data, that is, masking out noise 
spikes, instead of real objects.  This would alter the noise
properties of our data, and lead to systematic errors in our surface
photometry.  However, we need the low threshold to ensure that the 
low surface brightness outer regions 
of large, luminous galaxies are being properly masked.  

We deal with the ``over-masking'' problem by running SExtractor without
de-blending the various detections, that is, not assigning faint isolated 
objects as part of a much brighter object.  Then, we removed all
sources whose total magnitude was fainter than a cut-off
value.  We found the cut-off value in two different ways.  First, we
created the raw galaxy brightness distribution for both clusters by
selecting all objects with a stellarity index less than 0.5, where the 
stellarity index defines the likelihood that a source is or is not
extended through measurements of image moments by SExtractor's 
neural network \citep{sex1996}.  We then noted where 
the raw galaxy brightness distribution slope rapidly increased.  This
will indicate the onset of the noise spikes.  Second, we ran 
SExtractor on the mathematical inverse of each cluster image:
\begin{equation}
I_{inverse}(x,y) = -I_{image}(x,y)
\end{equation}
where x,y are the pixel coordinates of the image, and I(x,y) is the
flux in ADU at each point.  We then found the brightness distribution 
of negative noise spikes, which should provide an accurate measure of
the cut-off value, assuming that the noise is symmetrically distributed.   
The results of both of these tests are displayed in Figure 8, 
and are in good agreement with one another.  We set the cutoff 
magnitude to 23.8 for \a1413 and 23.0 for \mkw7, and removed all
sources from the segmentation image that were below this value.   

There is one other change that we must make to the segmentation
image.  After inspection of the corrected segmentation mask 
multiplied by the data, we occasionally found small groups of pixels 
that were completely surrounded by a large number of masked pixels.  
These ``islands'' of unmasked pixels are due to SExtractor treating
this small area as a separate object within the larger source. 
The islands  were corrected in the segmentation image by an automated 
process.  We masked each individual pixel that was surrounded by $N$ 
already masked pixels, and we repeated this 
process $M$ times.  By experimentation, we 
found that $N = 6$ and $M=20$ filled in the majority of the 
island-like structures with minimal changes to any other region.  
Finally, the images multiplied by the mask were visually inspected, 
and any regions that needed any further masking were masked using IMEDIT.  
These regions were mostly large-scale islands of unmasked pixels that
were not removed by our automated procedure.
Less than 2\% and 0.5\% of the pixels in \a1413 and \mkw7 respectively were 
removed manually.  The fraction of the images that was masked at 
this point is 43.4\% for \a1413, and 52.36\% for \mkw7.  
Figure 9 shows a sub-region of the \mkw7 image that contains stars 
and galaxies through each step of the masking process.

\subsection{Final Sky Subtraction, Masking and Large-scale Flat-Fielding 
Errors}

Accurate sky subtraction is crucial to determine the true amount of
intracluster starlight in each cluster, and is one of the dominant
sources of error in our analysis.  We now find a more accurate 
sky level for each cluster by using the masked image.  We first bin 
up the entire image into squares of 49 $\times$ 49 pixels.  
For each bin, we calculate a robust average \citep{sb1994}, 
ignoring all masked pixels; the results are displayed 
in Figures 10 and 11, respectively.  It is important to note that 
the entire greyscale range displayed in 
Figures 10 and 11 is $\pm$ 5 ADU from the sky level, which corresponds 
to a surface brightness of \muv = 26.7 \magsec, or 5.5 magnitudes 
below the sky level.

Several distinct features are apparent in these binned images.  First, 
the bright central portion of the cD is completely masked.  Then, as the 
distance from the cD increases, there is an annulus of bins where 
almost all pixels are masked, except for a few pixels that are 
significantly below the median value of the total number of 
pixels in the bin.  
This is the origin of the lower-flux ``ring'' seen around each cD galaxy
and is simply an artifact of masking the cD.  As we move even further
outwards, we find a region where the cD halo is still detectable, 
but it has dropped below the surface brightness at which SExtractor 
masks individual pixels (\muv $\sim$ 27.3 \magsec).  Finally, at the edges, 
the flux comes to a more or less constant value.  

However, in the case of \a1413, there is an additional low-surface 
brightness feature stretching along the right side of the frame
at an amplitude of $\sim 1$ ADU ($\sim$ \muv = 28.5 \magsec).  
This feature is almost certainly instrumental in nature, 
corresponding to the vignetting of the southern region of the
T2KA chip by the 2.1~m guide camera \citep{imaging2000}\footnote{this
document is available at http://www.noao.edu/kpno/kpno.html}.  This 
flat-fielding residual is small, but since it is systematic in 
nature, we chose to mask all pixels in the \a1413 image 
with x $>$ 1300.  For \mkw7, we see no evidence for this 
effect, and so we do not mask further in this case. 

To better determine our sky-values, and to measure our large-scale 
flat-fielding errors, we fit and subtract a plane from 
each masked, binned cluster image, using the IMSURFIT task in 
IRAF.  We took care to use regions 
on each image that are well away from the central cD.  The mean 
corrections from this step are small: less than 0.5 ADU for 
both clusters on average.  However, we emphasize that this process 
will remove any ICL that covers the entire image.  
We then created a histogram of sky values in 49 $\times$ 49 pixel 
bins well away from the center of each 
cluster.  We also required that the bins contain at least 200 
unmasked pixels to be included in the histogram.  There are 653 
such bins in \a1413 and 746 bins in \mkw7.    
These histograms are displayed in Figure 12.  The
width of the histograms provides a measure of our uncertainties due to
large-scale flat-fielding errors, and the faint outer wings of stars 
and galaxies that remain unmasked, even after the involved procedure
above.  We find that the large-scale flat-fielding error for both 
image is conservatively 1~ADU\footnote{equivalent to 1.3$\sigma$ if 
the distribution is Gaussian}, 
which corresponds to an uncertainty of 0.11\%. 

\subsection{Constructing the surface brightness profile}

We now un-mask the region around the central cD galaxy, and 
re-build the mask leaving the cD + intracluster light intact.  
We proceed as follows: we first use the ELLIPSE task in IRAF/STSDAS
\citep{busko1996}, based on the algorithms of \citet{jedr1987}, 
to obtain an approximate geometrical 
model of cD + intracluster light.  We then subtract this model 
from the data, and mask all the stars and galaxies that are 
superimposed over the cD using the same techniques as before.  
With this improved mask, we create a better model using ELLIPSE, 
and repeat the process until the residuals from the subtracted 
image are minimized.  This process was repeated seven times 
for each cluster.

In the case of \mkw7, a complication occurs at this step.  
There is a bright ($m_{v} \approx$~11.9) saturated star that lies 
within $21\parcsec1$ of the nucleus.  Naturally, we mask the inner
regions of this star, but because it is so bright, its radius
of influence extends over much of \mkw7's nucleus.  To remove
its influence on our surface photometry, we found its magnitude 
from a series of 10 second linearized exposures taken at the
same time as our surface brightness data.  We then subtracted 
the magnitude scaled PSF (\S 6.1) from the \mkw7 data.  This
subtraction is good to 0.05 magnitudes, and that error is incorporated 
into the error model for those bins.  Because the bright core of the
star is masked, the increase in the error is actually quite small.

We now bin up the un-masked data into regions whose size varies 
from a resolution element (5 $\times$ 5 pixels) 
near the cluster center, to the maximum
49$\times$49 pixel bin at the edges, using the robust mean as before.      
The binsize was increased exponentially with distance in the
x and y directions, so that the signal-to-noise ratio did not
strongly vary from the inner to outer regions.  The scale length was
100 pixels in each direction.  There are 11,037 such bins in 
the \a1413 image, and 7,925 bins in the \mkw7 image.  We next transform 
the mask-weighted x and y coordinates of each bin to the 
appropriate elliptical coordinates.  We do this by taking the best 
results from the ELLIPSE runs above, which consist of a table of 
the best-fitting elliptical isophotes as a function of semi-major axis 
\citep[for full details, see][]{jedr1987}  For each bin, we adopt 
the ellipticity and position angle from the nearest elliptical 
isophote from the ELLIPSE table.  In some cases, the ellipticity 
and position angles shifted abruptly in a non-physical manner.  
Therefore, we boxcar smoothed the ellipticity and position angle 
tables before applying them to our data.  With the x and y 
coordinates from our binning program, and the 
adopted, boxcar smoothed ellipticity and position angles 
from the ELLIPSE runs, we have now defined a unique ellipse for each
bin, with a semi-major axis $a$  and semi-minor axis $b$, with 
our bin at an eccentric angle $E$.  The surface brightness 
profiles for \a1413, and \mkw7 are displayed in Figure 13.  
In all cases, we define our radial coordinate $r$ 
as the geometric mean of the semi-major and semi-minor axes: 
$r = \sqrt{ab}$.  

\subsection{Limits to Our Precision}
The flux error model is described in detail in the 
Appendix.  To illustrate, we work through the errors in a 
5$\times$5 pixel bin, located 98 pixels in radius from the center
of \a1413.  In this bin, the mean number of counts is 40.9$\pm$1.9 
ADU above the sky level.  The errors are summarized in Table~1.  At 
large radii from the cD, the largest sources of error 
are large-scale flat-fielding errors, which are systematic, 
and do not depend on bin size.  Our errors at these large radii 
are $\sim 1.2$ ADU per bin.  Therefore, for a surface brightness 
bin to have a signal-to-noise ratio of at least five, it must have 
a mean value of at least 6 ADU, which corresponds to a surface 
brightness of \muv = 26.5 \magsec.  The signal-to-noise ratio approaches 
unity at \muv = 28.3 \magsec.
 
There is an additional source of error due to the coordinate
transformation via the ELLIPSE fits.  Any error in the ellipticity 
or position angle adopted will translate to an error in ellipsoidal
radius.  To quantify this error, we
propagated the error bars for the ellipticity and position angle 
derived from the ELLIPSE task through our transformation formulae.  
The error is typically 1.0\% in the radial direction.  Although
this error seems small, it does have a significant impact on the
errors for each bin.  If we transform the error in the radial 
coordinates to the corresponding error in magnitudes, we find 
the error is typically 0.04 magnitudes, assuming that the 
light follows a \rquart law.  To test the accuracy of the 
error bars from ELLIPSE, we simulated a series of 
images using tasks in the ELLIPSE package, and the parameters
of \mkw7's best-fitting model.  We then applied our error model 
to make noisy images from this model, re-ran ELLIPSE, and 
measured the dispersion in the measured parameters.  We found 
that the ELLIPSE task gave reasonable error estimates, actually 
overestimating the error by about a factor of two.  To be 
conservative, we adopt the ELLIPSE errors as they stand.

In short, we are confident that we have identified the major sources
of error due to instrumentational, observational, and computational 
sources.  A detailed, quantitative understanding of our errors is 
crucial for accurate measurement of the surface brightness profile.
One distinct advantage of this rigorous approach is that we can search
for non radial features in our data that a simple average would miss.

\subsection{Comparison with Published Results}

In order to directly compare our data against previously published results,
we first azimuthally average our surface brightness bins.  We then 
compare our \a1413 data to the $V$ photographic surface photometry of 
\citet{schombert1986} and our \mkw7 data to the $B$ photographic data of 
\citet{morbey1983}. The results are plotted in Figures 14 
and 15, respectively.

For \mkw7, after a displacement of 1.1 magnitudes to account 
for passband differences between the two images, and the average color
of the galaxy light, the difference between the two data sets 
is less than 0.1 mags everywhere but at 
very large radii.  The \citet{morbey1983} data is slightly brighter 
than our data at radii between 16 and 23 arcseconds; 
this may be due to the influence of the bright star 
21$\parcsec1$ away from \mkw7's nucleus.  Otherwise, the 
agreement is very good.  

For \a1413 however, our data do not agree as well with that of
\citet{schombert1986}.  There are clear systematic differences 
between the two radial profiles
at both large and small radii.  In the inner portions, 
the \citet{schombert1986} data are systematically
brighter by up to a magnitude.  This may be due to difficulties in
transforming measured photographic densities to magnitudes at higher flux
levels.  As evidence of this, we compare our CCD data at 
small radii to the CCD data of \citet{sgh1983}; which consists
of a Gunn $r$ radial surface brightness profile of \a1413, 
taken under similar seeing conditions  ($1\parcsec22$ versus 
$1\parcsec47$).  After a displacement of 0.3 magnitudes to account
for the cD's color (plotted in Figure 16), 
the two CCD data sets are in excellent agreement.  

At large radii, our measurements find systematically less 
flux than the \citet{schombert1986} data.  Figure 17 shows this 
region of discrepancy in more detail.  Unfortunately,
the area of comparison is exactly the region where our
signal-to-noise is rapidly decreasing and where sky-subtraction 
dominates the errors.  Figure 17 also shows the effect on our measured 
surface brightness profile if we had over-estimated the sky value 
by 1 ADU (if, for example, our field of view 
had not reached the sky).  As can be clearly seen, even a small 
error in our sky can alter the results dramatically in this case.  
Therefore, although we measure less flux than \citet{schombert1986} 
in this region, we cannot convincingly argue that our data is favored.  
Due to the large angular coverage of the photographic data, the sky
subtraction of \citet{schombert1986} may be more complete than our
own.

Could the discrepancy between our data and those of Schombert etal be
due to the presence of color gradients in the cD envelope? Since we
have not included color terms in our transformation from observed
Washington M to Johnson V, any underlying color gradient could
systematically affect our photometry and produce the observed
discrepancy between the two datasets. In practice, however, the effect
is small.  Mackie, Visvanathan, \& Carter(1990) have made a 
study of color gradients in central dominant galaxies, and they 
have found that the gradients are small, generally 
less than 0.2 magnitudes in (B-V) over the entire radial range 
observed.  Additionally, Mackie (1992) studied the colors of cD 
envelopes and found that their color profiles were also quite flat.  
Since we have calibrated our data for the mean (B-V) color of 
cD galaxies (B-V = 1.0), and given color term derived 
in \S5.4, this means that at most, 0.04 magnitudes of the 
offset can be attributed to color terms in our CCD data.  
Therefore, color terms cannot solely account for the discrepancy.  

\subsection{The Surface Brightness Profiles of \a1413 and \mkw7}

We now fit the surface brightness profiles of \a1413 and \mkw7 using
the \citet{sersic1968} profile:
\begin{equation}
I(r) = I_{e}~10^{-b_{n}[(r/r_{e})^{1/n}-1.]}
\end{equation}
where $b_{n}$ is a constant chosen so that half the total luminosity
predicted by the law is interior to $r_{e}$, and is well approximated
by the relation $b_{n} = 0.868n - 0.142$.  $I_{e}$ is the intensity
at the effective radius.  Because cD halos may exist as 
excesses above a best-fitting Sersic profile, we first fit the 
inner regions of the surface  brightness profiles (\muv $<$ 26.).  
Additionally, to ensure that our results are not affected by seeing, 
we ignore all data that has a radius less than three times the measured 
FWHM of each image.  We find that the best-fitting $n$ values for both
\a1413 and \mkw7 are indistinguishable from $n=4$, 
the de Vaucouleurs profile \citep{gdv1948}.  The best-fitting 
\rquart parameters for the inner regions of \a1413 and \mkw7
are given in Table~2.  We note that the reduced $\chi^{2}$ values for both
fits are quite high: 3.2 for \a1413 and 15.9 for \mkw7.  We discuss
the causes of this in the next section.  

\subsection{Deviations from the \rquart law:}

There are two possible explanations for why the fits have 
high reduced $\chi^{2}$ 
values: 1) we have underestimated our error bars substantially or 
2) there are real deviations in each cD galaxy from an \rquart law.  As 
we have stated earlier, we believe we have addressed all significant 
sources of error, including sky subtraction and other systematic sources.  
Therefore, we turn to possible deviations as the source of the 
large residuals.

On large angular scales, it is believed that excesses 
above the \rquart law for cD galaxies at large radii are due 
to what was classically called cD envelopes \citep{tonry1987,schombert1992}.  
We now search for the presence of such envelopes in our 
data.  After subtracting the best-fit \rquart law found above for 
each cluster, we fit another simple model to our data.  We assume 
that at a semi-major axis smaller than a radial scale $r_{cutoff}$, that 
there is no measurable excess in the surface brightness profile over 
the \rquart law.  At radii larger than $r_{cutoff}$, 
there is an excess above the best-fitting de Vaucouleurs model 
that is linear with \rquart, and has a slope $\beta$:

\begin{equation}
\mu = 0;~(r \le r_{cutoff});
\end{equation}  

\begin{equation}
\mu = \beta [r^{1/4} - r^{1/4}_{cutoff}];~(r > r_{cutoff})
\end{equation}

We emphasize that this parameterization is not intended to act as
a physical model, but rather a simple way of quantifying any
luminosity excess.  We now fit our data using standard 
least-squares methods to this model.  To obtain a robust result, 
we limit our fitting to where our data has a signal-to-noise 
of five or greater (\muv $<$ 26.5 \magsec).

The results are plotted in Figure 18, and the 
best-fitting parameters are given in Table 3, columns 1--5.  We find 
that both \a1413 and \mkw7 have clear excesses above the \rquart law.  
This is not unexpected for \a1413 \citep{schombert1986}, but the excess for 
\mkw7 is completely unexpected.  \citet{thuan1981} studied the 
surface brightness profile of nine brightest cluster members 
in MKW/AWM clusters and found that all of them followed an \rquart 
profile out to large radii.  Because of the lack of excess,
\citet{thuan1981} then argued that such brightest cluster galaxies 
were not ``true'' cD galaxies.  In their interpretation, 
cD envelopes are formed from galaxy collisions, and hence
will only be found in rich clusters. 

Here, we have found the exact opposite behavior: the poor 
cluster has a definite excess.
In the case of \a1413, we also find a clear excess, but it 
is significantly smaller than that found by \citet{schombert1986}.
The lower inferred excess is due to the discrepancies at both 
large and small radius between our data and those of Schombert
(1986).  The steeper inner profile of Schombert (1986) results
in a steep \rquart fit, enhancing the excess at large radius.
Our inner data points produce a shallower \rquart fit, and our 
outer data show lower surface brightnesses than Schombert (1986).  
Both effects significantly reduce the inferred luminosity excess 
over the single \rquart fit. 

To quantify these effects, we integrated our best-fitting
\rquart profiles, and our model for the excesses.  The results
are given in Table~3, columns 6--8.  We find that the fraction of
total luminosity in the excess component at the radius where our
data reaches a signal-to-noise of five is 13\% for \a1413, and 
21\% for \mkw7.  We then extrapolated our value for the 
\a1413 excess out to very large radii, in order to compare with 
the results of \citet{schombert1988}.  We stress that this 
extrapolation is very uncertain, as the value strongly depends on 
the accuracy of the slope $\beta$, and whether such a simple model 
is reasonable at very large radii.  We find that the fraction of excess 
luminosity to the total luminosity increases to 45\%.  However, we 
also find that the total derived luminosity is a factor of $\sim 2$ less, 
and the luminosity of the excess component is a factor of $\sim 3$ 
less than \citet{schombert1988} found.  

With the addition of the envelopes to our models of the 
surface brightness distribution in \a1413 and \mkw7, the reduced 
$\chi^{2}$ values show a large improvement: 2.1 for \a1413 
and 4.3 for \mkw7.  However, given that we have high confidence 
in our error models, such high reduced $\chi^{2}$ values are still 
unacceptable.  Therefore, there are still additional deviations 
from the elliptically symmetric flux model we have adopted.  

This result is supported by independent numerical results from 
the ELLIPSE program fits.  Numerous surface brightness studies 
of elliptical galaxies have shown that the 
surface brightness profiles do not follow perfect 
ellipses \citep[\eg][]{jedr1987,pel1990}.  These non-elliptical 
terms are often parameterized as the third and fourth-order terms of a 
Fourier series:
\begin{equation}
I(\theta) = I_{0} \left (\sum_{n=3}^{4} A_{n} sin(n\theta) + 
\sum_{n=3}^{4} B_{n} cos(n\theta) \right)    
\end{equation} where $I_{0}$ is the mean intensity of the 
elliptical isophote, and $\theta$ is the angle around the ellipse.      
The ELLIPSE program calculates these parameters automatically, and 
in Figure 19, we plot the derived A$_{3}$, B$_{3}$, A$_{4}$ and 
B$_{4}$ terms for \a1413 and \mkw7.   The terms are expressed as 
the Fourier amplitudes normalized by the semi-major axis and the 
isophotal gradient. The third order terms describe asymmetries in
the light profile, while the 4th order term -- in particular, A4 -- 
describes ``disky'' (negative A4) or ``boxy'' (positive A4) isophotes.
In both clusters, there are multiple regions where these terms
are significantly non-zero, with amplitudes ($\sim$ 5\%) much greater
than those customarily seen in normal ellipticals 
\citep[0.5\%;][]{jedr1987,pel1990}. 

In elliptical galaxies, these higher order isophotal coefficients are
often used to search for the presence of disks or the effects of
discrete mergers \citep[\eg][]{bender1989,rix1990}.  
In the distribution of ICL, which
extends out to hundreds of kpc, this interpretation needs to be
modified. To determine what causes these non-elliptical components in
our data, we subtract the best-fitting elliptical isophotal 
profiles, including both the
\rquart law and the excess component from both clusters, and then
examine the residuals for bins that are brighter than \muv = 26.5 \magsec.  
These residuals are displayed in Figures 20 and 21.  

In both cases, the elliptical fits break down near the center 
of the cD galaxy.  This is 
not unexpected: the number of unmasked pixels to fit 
are very few, and the ELLIPSE algorithms are known to systematically 
underestimate the ellipticity of galaxies in the very center 
\citep{jedr1987}.  In the case of \mkw7, the situation is 
particularly bad due to the bright star and galaxies very near 
the nucleus.  However, even if we exclude these inner regions, 
and a handful of bins that have clearly discordant flux, the 
$\chi^{2}$ values for the fits are still deviant: 1.49 for \a1413 
and 1.86 for \mkw7.  

In the case of \a1413, there appears to be a series of positive
and negative residuals, at a radial scale of $\sim 36\arcsec$.  
These residuals might be caused by a low surface brightness 
bridge between two large cluster galaxies present at that radius.  
Although there are tantalizing hints of diffuse tidal features 
in this residual image, none of them are clear enough to be 
definitive detections.  However, for \mkw7, there is a 
plume-like feature clearly 
seen in the residual image.  This feature has a mean surface 
brightness of \muv = $25.6 \pm 0.2$ \magsec, and is approximately 
61\arcsec~long ($\sim 32$ kpc).  It is approximately 32\arcsec ($\sim$ 17 
kpc) in width at the base of the plume, narrowing near the tip to 
$\sim 24\arcsec$ ($\sim$ 12 kpc).  The presence of this plume drives 
the ELLIPSE fits to generate a $\cos (n\theta)$ residual all the way 
around the ellipse, and is the main cause of the high reduced 
$\chi^{2}$ found for \mkw7.  If we approximate this plume as 
triangular in shape, we find a total magnitude of V$\sim$ 18.  
Given our adopted distance modulus to \mkw7, this is equal to the 
luminosity of a small galaxy ($M_{\mbox{v}} \sim -17$).  
This plume is clear evidence for ongoing tidal activity in \mkw7, and 
is similar in appearance to tidal debris found in the halo of M~87 
\citep{weil1997}.  

\subsection{Geometric properties of \a1413 and \mkw7}

From our adjusted fits from the ELLIPSE program, we also obtain 
the geometric parameters of the cD + intracluster light, such as the 
ellipticity and the position angle of the best-fitting ellipses.  
Those results are plotted in Figure 22.  
For \a1413, we have compared our results with those of 
\citet[][hereafter PSH]{porter1991}, and 
find good agreement over the range of radii we have in common.

The ellipticity of both clusters increases steadily with radius.
This is in good agreement with the results of PSH, 
who found an identical trend at smaller radii 
with a sample of 175 brightest cluster ellipticals.  However, 
the ellipticity ($\approx 0.8$) of \a1413 at large radii is 
extraordinary.  At a radii of 64~kpc (assuming H$_{0} = 
$60~km~s$^{-1}$~Mpc$^{-1}$, q$_{0} = $0.5), PSH 
found the average ellipticity of brightest cluster
ellipticals to be $\approx 0.4$, and the maximum ellipticity to be 
0.59, making \a1413 the most flattened brightest cluster galaxy 
ever measured.  However, the maximum value of PSH's 
ellipticity distribution {\it is the value for \a1413}.  It is 
therefore 
unclear whether the large ellipticity of \a1413 is peculiar to 
this cluster, or whether ellipticity for many brightest cluster 
galaxies continues to increase beyond the radial limits observed by 
PSH.  In contrast, the ellipticity 
distribution of \mkw7 is much more typical of that previously 
observed: a smooth rise to a maximum value of $\approx 0.4$.
This might be due to the fact that we probe a smaller range of
physical radii in \mkw7, compared to \a1413. 

In terms of position angle, \a1413 has only small isophote twists:
less than two degrees change overall.  \mkw7 has a large, but
not extraordinary twist of twenty degrees near the center, followed
by a gradual change in position angle out to large radius.  These
patterns are common for brightest cluster ellipticals 
\citep{porter1991}.

\section{The Search For Tidal arcs}

With the large scale properties of intracluster light + cD galaxy
established, we now focus on searching for smaller scale tidal debris 
arcs.  For the purposes of this search, we define an arc 
as an extremely elongated (ellipticity $\geq 0.5$) discrete 
object that can be detected visually.

We take the residual images constructed in \S6.7 and visually search them 
for the presence of any tidal debris arc structures.  We took two 
steps to ensure that residuals from the cD subtraction process were not 
mistakenly identified as tidal arcs.  First, we avoided the very 
inner 10\arcsec~radius of the center of each cluster, where the residuals
are the strongest.  Second, we also demanded that the arc candidate 
be visible in the unsubtracted cluster image, as well as the 
cD-subtracted image.  We found a total of five arc-like candidates 
in \a1413, and one candidate in \mkw7.  These arc-candidates are 
shown in Figures 23 and 24, respectively.  

The candidate arcs were then analyzed using the SExtractor 
software package.  
Astrometry for the central portion of the arcs was derived using the 
USNO-A 2.0 astrometric catalog \citep{monet1996,monet1998}, 
and the FINDER astrometric package from IRAF, and are accurate to 
0.3\arcsec.  The results of the analysis are given in Table~\ref{arcs1}.  
We compare these results to the tidal arcs previously found in the Coma and 
Centaurus clusters \citep{tren1998,gregg1998,cr2000}, whose properties 
are presented in Table~\ref{arcs2}.  Since these other observations are
taken in different filters than V, a color correction must be applied.  
\citet{gregg1998} found optical colors of 
B-V $\approx$ 0.9, V-R $\approx$ 0.6, V-I $\approx$ 1.2 for their 
debris arc candidate, typical of old stellar populations, so we adopt 
these colors for comparison purposes.  

We find that the arc candidates found in \a1413 and \mkw7 are
significantly shorter ($\sim 10$--$20$ kpc compared to 100 kpc), and
generally have higher surface brightness ( \muv $\sim 25.5$
mag~arcsec$^{-2}$ compared to 26 \magsec) than the tidal debris arcs
seen in Coma and Centaurus.  We conclude that, down to a limiting
surface brightness of \muv = 26.5 \magsec, there are no tidal debris arcs
longer than 30 kpc in either \a1413 or \mkw7. Given the depth of our
images, if either cluster contained long arcs such as those detected
in Coma and Centaurus, we would have detected them. The fact that we
do not detect them, particularly in a rich cluster like \a1413, argues
for real differences in the intrinsic ICL properties of massive
clusters.

So, what are these smaller arc-like objects that we do detect?  Arc
candidates 1--3 of Abell 1413 lie tangentially to the cD galaxy,
implying that these arcs may be due to strong gravitational lensing.
Gravitational arcs in clusters at this redshifts are uncommon
\citep{fort1994}, but some have been observed
\citep[\eg][]{campusano1998,blake1999}, and theoretical calculations
indicate that they should be detectable \citep{nata1997,cyp2001}.
Spectroscopic follow-up observations will be needed to prove whether
these arcs are gravitational in nature.  Arc candidates 4--5 of \a1413
may be other gravitational arcs, or genuine tidal debris.  For the
\mkw7 arc candidate, due to the lower redshift of the cluster, and the
fact that the \mkw7 arc candidate is extended in both dimensions, it
is unlikely that this arc is due to gravitational lensing. We may be
witnessing the early stages of the disruption of a dwarf cluster
galaxy, but without a redshift to assure cluster membership, this
conclusion is perhaps premature. While the \mkw7 arc candidate lies
within, and is perpendicular to the tidal plume detected there, this
is likely to be coincidental.

\section{Discussion and Summary}

We have performed deep surface photometry of two galaxy clusters
of greatly varying richness: \a1413 and \mkw7.  We find that both
galaxy clusters contain intracluster light out to large radii. The
cD envelopes of both clusters follow an \rquart profile over
a large range in radius, but also show an excess of diffuse light
at the largest radii. We also find evidence for substructure in
the ICL in \mkw7 in the form of a tidal plume and a single small
arc structure, and a set of small arc-like structures in 
\a1413 which may either be tidal in origin, or possibly due to 
gravitational lensing.

The accepted view of cD galaxies \citep{tonry1987,schombert1992} is 
that brightest cluster galaxies in rich clusters have large excesses 
in their surface brightness profiles over an \rquart law, and are
denoted as type cD, while brightest cluster galaxies in poor clusters 
do not have an excess, and are usually given a different designation 
(type D).  However, this view was established with photographic 
data, and newer CCD observations may cause this view to be revised.  
Multiple authors using CCDs have observed brightest cluster galaxies in 
rich clusters and have found that they follow an \rquart law out 
to very large radii \citep{uson1991b,scheick1994,gon2000}.  
In our particular case, we observed a classical ``cD envelope'' 
cluster (\a1413), and detected a much smaller envelope than the 
original photographic data found.  In addition, we found a clear 
excess in a poor cluster (\mkw7), where photographic data of similar
clusters has found no excess \citep{thuan1981}.  Although more observations
are needed, especially with clusters observed both with photographic
data and CCDs, it seems clear that the exact nature of cD envelopes 
needs to be re-evaluated.

If some rich clusters have ICL that follow 
\rquart profiles, while others do
not, it is possible that we can use the profiles to place constraints
on ICL formation mechanisms.  As mentioned in \S1, while violent
relaxation is known to produce a \rquart profile, tidal stripping may
produce a wider variety of profiles, depending on the distribution of
energy and angular momentum of the stripped population.  This would
imply that a cluster whose ICL followed an \rquart profile is
dynamically relaxed, and produced the majority of its ICL in the
process of cluster formation, when the gravitational potential is
rapidly changing.  In contrast, deviations from the \rquart law would
imply that intracluster star production is ongoing, and the cluster is
not dynamically relaxed. The fact that our two clusters, which are
quite disparate in mass, both show good \rquart profiles to large
radius favors models where the bulk of ICL is produced early during
cluster collapse. However, ongoing stripping does occur, as evidenced
by the tidal plume and small luminosity excesses in the very outer
regions of the clusters.


Separate from the radial profiles of ICL is the presence or absence of
tidal debris (plumes and arcs).  In our observations of \mkw7, we
have clearly found evidence for a tidal plume, much like that seen 
in M~87 by \citet{weil1997}.  The luminosity is small, like that of a
small spiral galaxy, but reinforces the finding that even for poor clusters, 
tidal stripping can be an important effect.

As for tidal debris arcs, we detected a number of possible short tidal
structures, but no long tidal arcs such as those seen in Coma
and Centaurus \citep{tren1998,gregg1998,cr2000}.  With a sample of 
only two galaxy clusters, it is premature to 
make any serious conclusions about the true frequency of long tidal 
debris arcs.  It might be that smaller-scale tidal structure in galaxy
clusters, such as that seen by \citet{conselice1999} are generally 
more common in galaxy clusters than long tidal debris arcs.  Long
tidal arcs are dynamically delicate, and may be  destroyed by
the passage of another galaxy in the cluster.  On the other hand, 
the lack of long tidal arcs might be due to the properties
of the clusters studied.  \a1413 appears to be dynamically evolved, and
perhaps tidal debris arcs are less common in such systems. \mkw7 is 
a much poorer cluster, so encounters are less common.  A larger 
sample of clusters is clearly needed for further progress.       

We do note that the vast majority of tidal debris seen in cluster 
simulations \citep{harass,dub1998,dub2000} has a surface brightness
much lower than our \muv = 26.5 \magsec~limit.  The structures that we have
observed so far may only be the brightest features in each cluster.  
Planned observations to deeper surface brightness limits will be important
in the future.

\acknowledgments
We thank P. Durrell for giving us the inverse image idea and 
we especially thank K. Stanek and B. Mochesjka for their help with
the T2KA linearity problem.  We thank the KPNO staff, especially 
R. Probst, for their help with the scattered light patterns.  We 
thank W. Pence for some early help, and his easy-to use 
CFITSIO routines, and we thank I. Busko and R. Jedrzejewski for 
their help with the ELLIPSE program.  We also thank an anonymous
referee for useful comments that improved the presentation of this
paper.

This work is supported by NSF through grants AST-9876143 (JCM) and
AST-9624542 (HLM) and by NASA through grant NAG5-7019 (JCM).  Funding
was also given by the Research Corporation's Cottrell Scholarships (JCM
and HLM).

\pagebreak
\appendix
\section{The Error Model}

It is necessary to have accurate error estimates of our surface 
photometry in order to perform the model fitting.  
Unlike earlier photographic work, deep CCD surface photometry
allows us to quantify measurement errors.  Measurement errors arise
from CCD behaviors such as readout noise and flat-fielding, as well from
sky noise.  Each error contribution will be addressed below.

\section{Readout Noise}
The readout noise per exposure is 1.1 ADU.  By combining 9 images for
\a1413, and 12 images for \mkw7 with a median, we are able to reduce the
effective read noise to
\begin{equation}
R_{eff} = 1.1~ADU~\frac{1.22}{\sqrt{N_{G}}} 
\end{equation}

The factor of 1.22 is due to the lower efficiency of a median over
a mean (see Morrison, Boroson, \& Harding 1994, Section 3.2.1).

\section{Photon Noise}
For $C$ ADU in a given pixel, the photon noise is (C/$g$)$^{1/2}$, where 
$g$ is the gain.  Combining 9 and 12 images respectively using a 
median reduces the photon noise to
\begin{equation}
\sigma_{Poisson} = \frac{1.22}{\sqrt{N}} \frac{\sqrt{C}}{\sqrt{g}}
\end{equation}

\section{Linearity Errors}
As mentioned in \S5.2 above, the T2KA chip has a known non-linearity.
We have corrected for this effect, but the parameters used for the
correction do not have infinite precision, and so our correction 
has errors.  The error in flux can be derived as follows:

\begin{equation}
\sigma_{linearity}^{2} = \sigma_{c1}^{2} C_{sky}^{2} + \sigma_{c2}^{2} 
\frac{C_{sky}^{4}}
{(32767)^{2}} + \sigma_{c3}^{2} \frac{C_{sky}^{6}} {(32767)^{4}}
\end{equation}
where $C_{sky}$ is the sky-subtracted flux.  Since this correction is small,
we apply it only to the flux, and not to any other calibration image. 
\section{Flat-Fielding Errors}
In principle, the only limit to the precision of the combined flat-field
images is the photon noise in the individual flat-field images.  This
small-scale variation is
\begin{equation}
\sigma_{sff} = \frac{\sqrt{C_{s}}}{\sqrt{g}} 
\frac{1.22}{\sqrt{N_{f}}} \frac{1.22}{\sqrt{N_{g}}}
\end{equation}
where $C_{s}$ is the number of counts in the final, combined master sky
flat image, g is the gain, $N_{f}$ is the number of individual sky flats
used to make the master sky flat, and N$_{g}$ is the number of individual
galaxy images used to make the final galaxy image.  The sky counts 
$C_{s} = 941$ ADU, the gain is 3.6 ADU, the number of sky images is 20,
and the number of galaxy images is 9 and 12, respectively.  The
percentage errors are 0.191\% for \a1413, and 0.166\% for \mkw7.  

In practice, the small-scale flat-fielding errors are not the only 
flat-fielding error we have.  There are also large-scale variations which
arise from the variation of the sky brightness across the image, 
instrumental effects such as flexure, and from the wings of bright
stars and galaxies that were not completely removed by combining
the individual sky flats.  Normally, to measure this effect, we prefer
to divide our sky flats into two sub-samples, create two sky flat
images from those sub-samples, and then find the standard deviation 
of the ratio of the two created flats.  However, we only have twenty
sky images, and dividing them up into two ten image sub-samples would be
too noisy for a realistic measurement. 

Instead, we masked each image, and constructed a histogram of sky
values (\S 6.2).  With 49 $\times$ 49 pixel bins, the noise between each bin
is completely dominated by large-scale flat-fielding errors, and 
the faint wings 
of unmasked objects.  We found an error of 1.0 ADU, which corresponds to
a fractional error of 1/941.0, or a percentage error of 0.11\%.   

\section{Surface Brightness Fluctuations}
For ultra-deep surface brightness observations of nearby galaxies,
a major source of error arises from intrinsic surface brightness
variations \citep{ts1988}.  For our distant galaxy clusters 
(see eq 10 of \citet{ts1988}), such an effect is completely 
negligible compared to our other errors.

\pagebreak

\pagebreak
\begin{deluxetable}{lll}
\tablewidth{0pt}
\tablenum{1}
\tablecaption{Errors in a 5 $\times$ 5 pixel bin with 40.9 ADU from \a1413
cD galaxy and 886 ADU from sky}
\tablehead{
\colhead{Source} 
& \colhead{Error in ADU} 
& \colhead{\% Error}
}
\startdata
Readout noise & 0.12 & 0.3 \\
Poisson Statistics & 1.5 & 3.7 \\
Linearity Error & 0.21 & 0.5 \\
Small-scale flat-fielding & 0.46 & 1.1 \\
Large-scale flat-fielding + Sky subtraction & 1.0 & 2.4 \\
\tableline
Total Error & 1.9 & 4.6 \\
\enddata
\end{deluxetable}

\pagebreak
\begin{deluxetable}{llllll}
\tablewidth{0pt}
\tablenum{2}
\tablecaption{Best-fitting de Vaucouleurs parameters}
\tablehead{
\colhead{Cluster} 
& \colhead{intercept of} 
& \colhead{slope of}
& \colhead{r$_{e}$} 
& \colhead{r$_{e}$} 
& \colhead{$\mu_{e}$}\\
& \colhead{best-fit (mag)} 
& \colhead{best-fit (mag/r$_{arcsec}^{1/4}$)}
& \colhead{(arcsec)} 
& \colhead{(kpc)} 
& \colhead{(mag / arcsec$^{2}$)}
}

\startdata
\a1413 & 16.63 & 3.71 & 25.4 & 57.4 & 25.0 \\
\mkw7  & 13.92 & 4.15 & 16.2 & 8.50 & 22.2\\
\enddata
\end{deluxetable}

\begin{deluxetable}{llllllll}
\tablewidth{0pt}
\tablenum{3}
\tablecaption{Residual model}
\tablehead{
\colhead{Cluster} 
& \colhead{r$_{cutoff}$} 
& \colhead{r$_{cutoff}$} 
& \colhead{$\beta$} 
& \colhead{$\beta$} 
& \colhead{m$_{\mbox{total, v}}$}
& \colhead{m$_{\mbox{excess, v}}$}
& \colhead{$\frac{L_{\mbox{excess}}}{L_{\mbox{total}}}$}
\\
& \colhead{(arcsec)} 
& \colhead{(kpc)} 
& \colhead{(mags r$_{arcsec}^{-1/4}$)}
& \colhead{(mags r$_{kpc}^{-1/4}$)}
& \colhead{(mag)}
& \colhead{(mag)}
& \colhead{}
}
\startdata
\a1413 & 29.2 & 66.0 & -1.11 & -1.01 & 14.7 & 16.9 & 0.13\\
\mkw7  & 22.6 & 11.8 & -1.24 & -0.78 & 12.6 & 14.3 & 0.21\\
\enddata
\end{deluxetable}

\pagebreak
\begin{deluxetable}{lllllllll}
\tablewidth{0pt}
\tablenum{4}
\tabletypesize{\footnotesize} 
\tablecaption{Properties of arc candidates in \a1413 and \mkw7\label{arcs1}}
\tablehead{
\colhead{Name} 
& \colhead{$\alpha$ (2000)} 
& \colhead{$\delta$ (2000)} 
& \colhead{Linear size}
& \colhead{Linear size}
& \colhead{m$_{iso}$} 
& \colhead{M$_{iso}$} 
& \colhead{$\mu_{max}$} 
& \colhead{$\mu_{avg}$} 
\\
& \colhead{}
& \colhead{}
& \colhead{(arcsec)}
& \colhead{(kpc)}
& \colhead{}
& \colhead{} 
& \colhead{} 
& \colhead{}
}

\startdata
\a1413 Arc~1 & 11h 55m 17.888s & +23\degr 24\arcmin 45.70\arcsec 
& 11.3 $\times$ 1.2\tablenotemark{a} 
& 25.4 $\times$ 2.8 & 22.6 $\pm$ 0.05\tablenotemark{b} 
& -16.3  & 24.8 & 25.4 \\

\a1413 Arc~2 & 11h 55m 20.326s & +23\degr 23\arcmin 19.94\arcsec 
& 7.0 $\times$ 1.2\tablenotemark{a}  
& 15.8 $\times$ 2.8 & 23.6 $\pm$ 0.1 & -15.3 & 25.1 & 25.9  \\

\a1413 Arc~3 & 11h 55m 18.372s & +23\degr 23\arcmin 57.34\arcsec 
& 6.1 $\times$ 2.0 & 13.7 $\times$ 4.5  & 22.7 $\pm$ 0.02 & -16.2  
& 25.0 & 25.4 \\

\a1413 Arc~4 & 11h 55m 19.193s & +23\degr 24\arcmin 26.80\arcsec 
& 3.9 $\times$ 1.2\tablenotemark{a} 
& 8.8 $\times$ 2.8 & 23.7 $\pm$ 0.08 & -15.2 & 25.2 & 25.4   \\

\a1413 Arc~5 & 11h 55m 19.667s & +23\degr 24\arcmin 24.51\arcsec 
& 6.0 $\times$ 1.2\tablenotemark{a}  
&  13.5 $\times$ 2.8 & 24.5 $\pm$ 0.14  & -14.4 & 25.5 & 26.6   \\
\tableline
\mkw7 Arc~1 & 14h 33m 58.084s & +03\degr 45\arcmin 57.90\arcsec 
& 4.9 $\times$ 1.8  & 2.6 $\times$ 1.0 
& 23.2 $\pm$ 0.08 & -12.1 & 25.0 & 25.8  \\
\enddata
\tablenotetext{a}{This dimension is unresolved in $1\parcsec22$ seeing}
\tablenotetext{b}{The magnitude errors are derived from SExtractor, 
and are an underestimate to the true errors.}
\end{deluxetable}

\begin{deluxetable}{llllllllll}
\tablewidth{0pt}
\tablenum{5}
\tabletypesize{\footnotesize} 
\tablecaption{Properties of previously discovered cluster tidal 
debris\label{arcs2}}
\tablehead{
\colhead{Cluster} 
& \colhead{Linear size}
& \colhead{Linear size}
& \colhead{m} 
& \colhead{M\tablenotemark{a}} 
& \colhead{$\mu_{max}$} 
& \colhead{$\mu_{avg}$} 
& \colhead{$\mu_{V}$\tablenotemark{b}}
& \colhead{Filter} 
& \colhead{Source}
\\ 
& \colhead{(arcsec)}
& \colhead{(kpc)\tablenotemark{a}}
& \colhead{}
& \colhead{} 
& \colhead{} 
& \colhead{}
& \colhead{}
& \colhead{}
& \colhead{}
} 

\startdata

Coma TM & 180 $\times$ 10 & 80 $\times$ 4 & \nodata & \nodata 
& 26.5 & \nodata & 25.6 & B & TM1998\\ 

Coma TM & & & \nodata & \nodata & 25.0 & \nodata & & R & TM1998\\ 

Coma LSB1 & 270 $\times$ 60 & 120 $\times$ 30 & 15.6 $\pm$ 0.1 
& -19.2\tablenotemark{a} $\pm$ 0.1 & \nodata & 25.7 &  26.3 & R & GW1998\\

Centaurus CR & 720 $\times$ 10 & 160 $\times$ 1 & 18.4 $\pm$ 0.5 & -13.1  
$\pm$ 0.5 & \nodata & 27.8 & 26.9 & B & CR2000\\

Centaurus CR & &  & 16.7 & -14.8 & \nodata & 26.1  & & R & CR2000\\

Centaurus CR & & & 16.4 & -15.11 & \nodata & 25.7 & & I & CR2000\\
\enddata
\vfill
\tablenotetext{a}{We assume a distance modulus of 34.83 to Coma, and 31.51
to Centaurus}
\tablenotetext{b}{Assuming B-V = 0.9, V-R = 0.6, V-I=1.2 \citep{gregg1998}}
\end{deluxetable}

\clearpage
\pagebreak
\begin{figure}
\figurenum{1}
\caption{The spectrum of the night sky at Kitt Peak National Observatory,
taken from the data of Massey \& Foltz (2000).  Overlaid over the spectrum 
is the filter transmission curve of the Washington {\it M} filter
used in these observations (KP1581), shown as the solid line.  For
comparison, a standard Harris V filter (KP1542) is also shown as the dashed
line.  The Washington M filter contains fewer strong sky-emission lines 
compared to the V filter, most notably O~I $\lambda$ 5577.}
\end{figure}

\pagebreak
\begin{figure}
\figurenum{2}
\caption{A comparison of T2KA photometry obtained from a 60s test 
exposure and a median-combined 900s exposure of \mkw7.  Although there is
large scatter (primarily from photon noise in the 60s exposure), there
is a clear systematic residual with instrumental magnitude, identical
to that seen by Mochesjka \etal (2001).  The line denotes the least-squares
linear fit through the data.}
\end{figure}

\pagebreak
\begin{figure}
\figurenum{3}
\caption{The difference between surface brightness of data corrected for the
non-linearity effect found by Mochesjka \etal (2001) for the T2KA
detector, and un-corrected data, over the range of surface 
brightnesses applicable to our survey.  Note that the effect is 
nowhere greater than 0.012 magnitudes, and the effect is relatively 
constant over the range of surface brightnesses that ICL would 
be present ($\mu_{V}$ = 24--30)}
\end{figure}

\pagebreak
\begin{figure}
\figurenum{4}
\caption{Images of the 20 blank sky frames used in constructing the master sky
flat, after being flat-fielded and 
averaged into 50 pixel bins.  The small scale variations are due to 
bright stars and galaxies in each individual frame, but the large scale
variations are due to changes in the sky illumination.  These
large-scale features are partially removed by the plane normalization.  
The grid-like scattered light pattern can also be barely seen in 
three of the frames (bottom row, first and fourth column, and 
second row from bottom, fifth column).  These grid-like patterns 
are almost invisible in the un-binned images.}
\end{figure}

\pagebreak
\begin{figure}
\figurenum{5}
\caption{Our final, median-combined image for \a1413.  North is at the left
of this image, and east is at the bottom.  The image is $10\parcmin2$
square, corresponding to a linear distance of 1.5 Mpc at our adopted 
distance to \a1413.  Note the incredible richness of this cluster: almost every
bright object in the frame with the exception of the two saturated stars
at the bottom is a galaxy.}
\end{figure}

\pagebreak
\begin{figure}
\figurenum{6}
\caption[]
{Our final, median-combining image for \mkw7.  North is at the left
of this image, and east is at the bottom.  The image is $10\parcmin2$
square, corresponding to a linear distance of 330 kpc at our adopted 
distance to \mkw7.  The bright, saturated star superimposed near 
MKW7s nucleus cannot be seen in this image greyscale.}
\end{figure}

\pagebreak
\begin{figure}
\figurenum{7}
\caption{The surface brightness profile of a saturated star on our \a1413
image, averaged azimuthally.  As can be clearly seen, the profile
extends to very large radii.  The solid horizontal line is set at the
surface brightness limit of 1 ADU~pixel$^{-1}$ above the sky value,
corresponding to a surface brightness of 28.48 \magsec.
}
\end{figure}

\pagebreak
\begin{figure}
\figurenum{8}
\caption{The brightness distribution of all objects in the 
\a1413 (left) and \mkw7 (right) fields found by SExtractor
that have a stellarity index less than 0.5, shown as the filled
dots. Note that this is not the cluster luminosity function, as
no background subtraction has been done, and the blending parameter of
SExtractor has been turned off.  The brightness distribution is
compared to the brightness distribution of negative noise spikes, shown as
the open diamonds.  As can be clearly seen, the brightness distribution
steepens at the same magnitude (m $\approx 23$) that the negative 
noise spikes become present in great numbers.  This denotes where 
the overmasking problem begins.}

\end{figure}

\pagebreak
\begin{figure}
\figurenum{9}
\caption{A region of \mkw7 shown through all stages of the masking
process.  From left to right, and top to bottom the sub-images are: 1)
The original image, 2) the image multiplied by the stellar mask found
through DAOFIND 3) the previous image multiplied by the mask
from SExtractor, 4) the previous mask, with the ``over-masking''
correction applied, 5) the previous image, with the surrounding
pixels correction made, and 6) the previous image after manual 
masking.  The grey-scale in all of the images is 10 ADU above and 
below the sky value, and the standard deviation of the sky 
background is 5.8 ADU per pixel.  The masking procedure removes the 
vast majority of stellar and galaxy light in the image.
}
\end{figure}

\pagebreak
\begin{figure}
\figurenum{10}
\caption{The binned-up image of \a1413, with a greyscale stretch of five 
ADU above and below the sky value.  North is again to the left, and east
is at the bottom of this image.  As the bins increase in radius from
the central cD galaxy, their fluxes systematically change, as
discussed in the text.  At the far right of this image, a large-scale
flat-fielding error of 1 ADU is clearly visible.}
\end{figure}

\pagebreak
\begin{figure}
\figurenum{11}
\caption{The binned-up image of \mkw7, similar to that of Figure 10.
In this case, no large-scale flat-fielding error is present.}
\end{figure}

\pagebreak
\begin{figure}
\figurenum{12}
\caption{The histogram of sky values for \a1413 (left) and \mkw7 (right), 
binned up into 0.25 ADU intervals.  See the text for the description 
of how this histogram was created. Ideally, the sky values should 
all equal zero, but due to 
large-scale flat-fielding errors, and the wings of unmasked stars and
galaxies, there is a dispersion about zero.  We estimate the error 
in the sky to be $\pm$ 1 ADU.}
\end{figure}

\pagebreak
\begin{figure}
\figurenum{13}
\caption{The radial surface brightness profiles of \a1413 (left) and 
\mkw7 (right), as measured by our data.  The solid horizontal solid 
line at $\mu \sim 26.5$ and the dashed
line at $\mu \sim 28.3$ indicate where our data has a signal-to-noise of
five and one, respectively.  The vertical dot-dashed line indicates the
radial scale of the seeing disk in these units.  Note that the effects of
seeing extend over several seeing radii.}
\end{figure}
\clearpage

\pagebreak
\begin{figure}
\figurenum{14}
\caption{The radial surface brightness profile of \mkw7, as measured by
our data (filled circles), compared to the $B$ band photographic 
surface photometry of Morbey \& Morris (1983) (open diamonds).  The
$B$ data has been shifted by 1.1 magnitudes vertically to account for
passband differences.  The dashed line at $\mu = 26.5$ indicates where our
data reaches a signal-to-noise of five and the solid line at $\mu = 28.3$
indicates where our data has a signal-to-noise of one.  The two profiles
agree to with $\pm$ 0.1 magnitudes except at large radii.}
\end{figure}

\pagebreak
\begin{figure}
\figurenum{15}
\caption{The radial surface brightness profile of \a1413, as measured by
our data (filled circles), compared to the $V$ band photographic 
surface photometry of Schombert (1986) (open diamonds). 
The dashed line at $\mu = 26.5$ indicates where our data reaches a 
signal-to-noise of five and the solid line at $\mu = 28.3$
indicates where our data has a signal-to-noise of one.  The two profiles
disagree at large and small radii.  See the text for discussion.}
\end{figure}

\pagebreak
\begin{figure}
\figurenum{16}
\caption{An expanded view of the surface brightness profile of \a1413 at small 
radii.  As before, the filled circles are our
data, but the open diamonds are now the $r$ band CCD data of 
Schneider, Gunn \& Hoessel (1983), after being displaced by 0.3 magnitudes.  
The two profiles are in good agreement.}
\end{figure}

\pagebreak
\begin{figure}
\figurenum{17}
\caption{A expanded view of the surface brightness profile of \a1413 
at large radii.  As before, the filled circles are our
data, the open diamonds are the data of Schombert (1986), and the
two horizontal lines show signal-to-noise ratios of five, and one
respectively.  The filled squares indicate the effect of overestimating
the sky by 1~ADU in our data.}
\end{figure}

\pagebreak
\begin{figure}
\figurenum{18}
\caption{The residuals in our surface brightness profile for \a1413 (left) 
and \mkw7 (right), after
the best-fitting de Vaucouleurs law of the inner regions has been
subtracted.  Negative residuals indicate a magnitude excess, and positive
residuals indicate a flux deficit.  The best-fitting excess
model (see the text), is plotted as the solid line.  There is a clear
deviation above the de Vaucouleurs law in both cases.  Such behavior for 
brightest cluster members in poor clusters is unexpected 
\citep{thuan1981}.}
\end{figure}

\pagebreak
\begin{figure}
\figurenum{19}
\caption{A plot of the A$_{3}$, A$_{4}$, B$_{3}$, and B$_{4}$ terms for 
\a1413 (left) and \mkw7 (right), as a function of radius. Clear 
non-zero terms are present in both data sets, indicating that there
are non-elliptical residuals present in the data.}
\end{figure}

\pagebreak
\begin{figure}
\figurenum{20}
\caption{Our residual image for \a1413, after the best-fitting
elliptical model of the cD + ICL has been subtracted.  The 
black ellipse shows where the measured surface brightness has
a signal-to-noise greater than five.  There is an indication of
a low surface brightness bridge between two luminous galaxies located
up and to the right of the cD nucleus (shown by the square), 
but the results are not conclusive.  No obvious large-scale 
tidal features are apparent, but there are a number of small arc-like
structures clearly visible.  See the text for discussion of these arcs.}
\end{figure}

\pagebreak
\begin{figure}
\figurenum{21}
\caption{Our residual image for \mkw7, after the best-fitting
elliptical model of the cD + ICL has been subtracted.  The 
black ellipse shows where the measured surface brightness has
a signal-to-noise greater than five.  A large tidal plume is apparent
leading from the center of the image to the right (south), 
and up (west) of the galaxy's nucleus.}
\end{figure}

\pagebreak
\begin{figure}
\figurenum{22}

\caption{The best-fitting values for the position angle and ellipticity
of our ELLIPSE fits, as a function of radius for \a1413 (top) and
\mkw7 (bottom).}
\end{figure}

\pagebreak
\begin{figure}
\figurenum{23}
\caption{An $2\parcmin7 \times 2\parcmin7$
image of the central region of Abell 1413, with the cD
galaxy subtracted, and point sources masked out.  North is to 
the left, and east is at the bottom of this figure.
Two arc-like structures are clearly visible (1, 2), with another 
three (3-5) possible.  Note that arcs 1-3 lie tangentially to the
cD galaxy, implying that these features might be due to gravitational
lensing.}
\end{figure}

\pagebreak
\begin{figure}
\figurenum{24}
\caption{An $1\parcmin4 \times 1\parcmin4$ image of the central 
region of \mkw7, with the cD galaxy subtracted.  North is to the 
left, and east is at the bottom of this figure.
There is one arc-like feature clearly seen away 
from the galaxy nucleus.}
\end{figure}

\end{document}